\documentclass[preprint,prd,aps,showpacs,showkeys,nofootinbib]{revtex4}
\usepackage{amssymb}
\usepackage{}
\usepackage{amsmath}
\usepackage{graphicx}
\usepackage{float}
\usepackage{multirow}

\begin{document}

\preprint{}

\title{Rare B meson decays in the Minimal R-symmetric Supersymmetric Standard Model}

\author{Ke-Sheng Sun$^a$\footnote{sunkesheng@126.com;\;sunkesheng@bdu.edu.cn}, Kui-Wen Guan$^a$\footnote{guankuiwen@bdu.edu.cn}, Hao-Yi Liu$^{b,c,d}$\footnote{3338183428@qq.com}, Jin-Lei Yang$^{b,c,d}$\footnote{jlyang@hbu.edu.cn},Tie-Jun Gao$^{e}$\footnote{tjgao@xidian.edu.cn}}

\affiliation{$^a$Department of Physics, Baoding University, Baoding 071000,China\\
$^b$Department of Physics, Hebei University, Baoding 071002, China\\
$^c$Hebei Key Laboratory of High-precision Computation and Application of Quantum Field Theory, Baoding, 071002, China\\
$^d$Hebei Research Center of the Basic Discipline for Computational Physics, Baoding, 071002, China\\
$^e$School of Physics, Xidian University, Xi'an 710071, China}

\begin{abstract}
Taking into account the constraints imposed by experimental data on the parameter space, we analyze the lepton flavor violating decays of B meson in the scenario of the minimal R-symmetric supersymmetric standard model. The prediction of the branching ratios is strongly affected by $\tan\beta$ and the off-diagonal entries in the slepton and squark mass matrices. The off-diagonal entries in the slepton mass matrix are constrained by the experimental limits of radiative two body decays of leptons.
The off-diagonal entries in the squark mass matrix are constrained by the experimental limits of low energy observables related to B meson physics. The branching ratio of $B^0_d\rightarrow \mu \tau$ is predicted to be four orders of magnitude below the future experimental sensitivity and the decay $B^0_d\rightarrow \mu \tau$ has a higher chance of being observed in the future.

\end{abstract}



\maketitle

\section{Introduction}
\indent\indent

The Lepton Flavor Violating (LFV) decays are of great importance in searching for new physics beyond the Standard Model (SM) since they are highly suppressed in the SM. Search for such LFV decays has been pursued to date in a host of processes of leptons, $Z$ boson, Higgs boson and various hadrons.  Examples of the LFV decay of B-hadron are $B^0_{d}\rightarrow l_1l_2$ and $B^0_{s}\rightarrow l_1l_2$, where $l_1\in (e,\mu)$, $l_2 \in (\mu,\tau)$ and $l_1 \ne l_2$. The present upper bounds on the branching ratio (BR) of $B^0_q\rightarrow l_1l_2$ (from now on, we use $B^0_q$ to denote both $B^0_d$ and $B^0_s$) from both the Belle data and the LHCb data are shown in TABLE.\ref{bsdexp} \cite{PDG}. Because of the extremely small neutrino masses, the Feynman diagram contributes a factor of at least $10^{-52}$ to the amplitude \cite{Bilenky,Esteban}, and therefore the branching ratio is far below any experimental sensitivity at present. Several overviews of both the theoretical motivations for charged LFV and the experimental approaches are given in Refs.\cite{Calibbi,Ardu,Lindner}.
\begin{table}[h]
\caption{Current and future limits on BR($B^0_q\rightarrow l_1l_2$).}
{\begin{tabular}{@{}c|c|c|c|c|c@{}}
\toprule
Decay &Current limit &Future limit&Decay &Current limit&Future limit\\
\colrule
$B^0_{d}\rightarrow e \mu $&$1.0\times 10^{-9}$&$9\times 10^{-11}$\cite{Aaij}&$B^0_{s}\rightarrow e \mu $&$5.4\times 10^{-9}$&$9\times 10^{-11}$\cite{Aaij}\\
$B^0_{d}\rightarrow e \tau $&$ 1.6\times 10^{-5}$&-&$B^0_{s}\rightarrow e \tau $&$1.4\times 10^{-3}$&-\\
$B^0_{d}\rightarrow \mu \tau $&$1.4\times 10^{-5}$&$1.3\times 10^{-6}$\cite{Altmannshofer}&$B^0_{s}\rightarrow \mu \tau $&$4.2\times 10^{-5}$&-\\ \botrule
\end{tabular}
\label{bsdexp}}
\end{table}

The LFV processes of B meson are associated with the lepton nonuniversality effect in semileptonic decays and $b\rightarrow sll$ transitions. These processes have been similarly explored in several new physical models, such as supersymmetric models \cite{susy,susy1,susy2}, models extended with extra gauge $Z^{'}$ boson \cite{Zp3}, heavy singlet Dirac neutrinos \cite{HDn}, leptoquarks \cite{Lq,Lq1,Lq2,Lq3}, the Pati-Salam model \cite{PS} and a minimal extension of the SM with one neutral singlet scalar\cite{ss1}. Full one loop calculation of BR($B^0_q\rightarrow l_1l_2$) is available in high energy physics softwares \cite{Dreiner}. In these references, the prediction of BR($B^0_{q}\rightarrow e\mu$) can be greatly enhanced, even up to $10^{-11}$, which are very promising detected in the future. BR($B^0_{q}\rightarrow e\tau$) and BR($B^0_{q}\rightarrow \mu\tau$) can also be enhanced close to $B^0_{q}\rightarrow e\mu$ \cite{Lq,Lq1}. 

In this paper, we investgate the LFV decays of B meson in the Minimal R-symmetric Supersymmetric Standard Model (MRSSM) \cite{Kribs}. The MRSSM is an extension of the Minimal Supersymmetric Standard Model (MSSM) that incorporates a global R-symmetry---a continuous U(1) symmetry acting on superfields \cite{Fayet,Salam}. The R-symmetry forbids many problematic soft-breaking terms, including the trilinear A-terms responsible for CP violation and flavor issues. The MRSSM introduces Dirac masses for gauginos by pairing the usual gauge superfields with additional chiral adjoint superfields. This contrasts with the Majorana gaugino masses in the MSSM and leads to distinct phenomenology \cite{Die1, Die2, Die3, Die4, Die5, Kumar, Blechman, Kribs1, Frugiuele, Jan, Chakraborty, Braathen, Athron, Alvarado,Jan2,ss2,ss3,ss4,ss5}. 

Taking into account constraints from electroweak precision observables, including W boson mass, oblique parameters and B meson decays, we show that the MRSSM can accommodate the observed 125 GeV Higgs boson as the lightest scalar of the model. The allowed parameter space is given at 1$\sigma$ level. Based on this parameter space and taking into account constraints from radiative decay $l_2\rightarrow l_1\gamma$, we give the upper predictions on the LFV decays of B meson. Similar to the case in the MSSM, the LFV decays of B meson mainly originate from the off-diagonal entries in slepton mass matrices $m_l^2$ and $m_r^2$. We also explore the LFV decays of B meson as a function of several model parameters. 

The paper is organized as follows. In Section \ref{sec2}, we provide a brief introduction to the model, and give the analytic expressions for every Feynman diagram contributing to LFV decays of B meson in detail. The numerical results are presented in Section \ref{sec3}, and the conclusion is drawn in Section \ref{sec4}.

\section{Formalism\label{sec2}}

In this section, we provide a simple overview of the MRSSM to fix the notations that will be used in the rest of the work. The MRSSM has the same gauge group $SU(3)_C\times SU(2)_L\times U(1)_Y$ as the SM and MSSM. 
Besides the standard MSSM matter, the spectrum of fields in the MRSSM contains Higgs and gauge superfields added by the chiral adjoints $\hat{\cal O},\hat{T},\hat{S}$ and two $R$-Higgs iso-doublets $\hat{R}_{u}$ and $\hat{R}_{d}$. The superfields and the component fields in the MRSSM are listed in TABLE.\ref{field}.
\begin{table}[h]
\caption{The R-charges of the superfields and the corresponding bosonic and fermionic components in the MRSSM. }
\begin{tabular}{@{}c|c|c|c|c|c|c@{}} \toprule
Field&Superfield&R-charge&Boson&R-charge&Fermion&R-charge\\
\colrule
Gauge vector & $\hat{g},\hat{W},\hat{B}$&0& $g,W,B$&0& $\tilde{g},\tilde{W}\tilde{B}$&  $+$1 \\\hline
\multirow{2}*{Matter}& $\hat{l}, \hat{e}$& $+$1& $\tilde{l},\tilde{e}^*_R$&$+$1& $l,e^*_R$& 0   \\
 \cline{2-7}     & $\hat{q},{\hat{d}},{\hat{u}}$& $+$1& $\tilde{q},{\tilde{d}}^*_R,{\tilde{u}}^*_R$ & $+$1& $q,d^*_R,u^*_R$                             & 0 \\\hline
H-Higgs & ${\hat{H}}_{d,u}$&0& $H_{d,u}$& 0& ${\tilde{H}}_{d,u}$&$-$1 \\ \hline
R-Higgs & ${\hat{R}}_{d,u}$ & $+$2& $R_{d,u}$&$+$2& ${\tilde{R}}_{d,u}$& $+$1 \\\hline
Adjoint chiral& $\hat{\cal O},\hat{T},\hat{S}$&0& $O,T,S$&0& $\tilde{O},\tilde{T},\tilde{S}$ &$-$1 \\\botrule
\end{tabular}
\label{field}
\end{table}

The general form of the superpotential of the MRSSM is given by \cite{Die1},
\begin{equation}
\begin{split}
\mathcal{W}_{MRSSM} &= \mu_d(\hat{R}_dH_d)+\mu_u(\hat{R}_uH_u)+\Lambda_d(\hat{R}_d\hat{T})H_d+\Lambda_u(\hat{R}_u\hat{T})H_u\\
&+\lambda_d\hat{S}(\hat{R}_dH_d)+\lambda_u\hat{S}(\hat{R}_uH_u)-Y_d\bar{D}(QH_d)-Y_e\bar{E}(LH_d)+Y_u\bar{U}(QH_u),
\end{split}
\end{equation}
where $H_u$ and $H_d$ are the MSSM-like Higgs weak iso-doublets, $\hat{R}_u$ and $\hat{R}_d$ are the $R$-charged Higgs $SU(2)_L$ doublets and the corresponding Dirac higgsino mass parameters are denoted as $\mu_u$ and $\mu_d$. $\lambda_u$, $\lambda_d$, $\Lambda_u$ and $\Lambda_d$ are parameters of Yukawa-like trilinear terms involving the singlet $\hat{S}$ and the triplet $\hat{T}$, which is given by
\begin{equation}
\hat{T} = \left(
\begin{array}{cc}
\hat{T}^0/\sqrt{2} &\hat{T}^+ \\
\hat{T}^-  &-\hat{T}^0/\sqrt{2}\end{array}
\right).
 \end{equation}
The soft-breaking scalar mass terms are given by
\begin{equation}
\begin{split}
V_{SB,S} &= m^2_{H_d}(|H^0_d|^2+|H^{-}_d|^2)+m^2_{H_u}(|H^0_u|^2+|H^{+}_u|^2)+(B_{\mu}(H^-_dH^+_u-H^0_dH^0_u)+h.c.)\\
&+m^2_{R_d}(|R^0_d|^2+|R^{+}_d|^2)+m^2_{R_u}(|R^0_u|^2+|R^{-}_u|^2)+m^2_T(|T^0|^2+|T^-|^2+|T^+|^2)\\
&+m^2_S|S|^2+ m^2_O|O^2|+\tilde{d}^*_{L,i} m_{\tilde{q},{i j}}^{2} \tilde{d}_{L,j} +\tilde{d}^*_{R,i} m_{\tilde{d},{i j}}^{2} \tilde{d}_{R,j}+\tilde{u}^*_{L,i}  m_{\tilde{q},{i j}}^{2} \tilde{u}_{L,j}\\
&+\tilde{u}^*_{R,i}  m_{\tilde{u},{i j}}^{2} \tilde{u}_{R,j}+\tilde{e}^*_{L,i} m_{\tilde{l},{i j}}^{2} \tilde{e}_{L,j}+\tilde{e}^*_{R,{i}} m_{\tilde{r},{i j}}^{2} \tilde{e}_{R,{j}} +\tilde{\nu}^*_{L,i} m_{\tilde{l},{i j}}^{2} \tilde{\nu}_{L,j}
\end{split}
\end{equation}
All trilinear scalar couplings involving Higgs bosons to squarks and sleptons are forbidden due to the $R$-symmetry. The soft-breaking Dirac mass terms of the singlet $\hat{S}$, triplet $\hat{T}$ and octet $\hat{O}$ take the form,
\begin{equation}
V_{SB,DG}=M^B_D\tilde{B}\tilde{S}+M^W_D\tilde{W}^a\tilde{T}^a+M^O_D\tilde{g}\tilde{O}+h.c.,
\label{}
\end{equation}
where $\tilde{B}$, $\tilde{W}$ and $\tilde{g}$ are usually MSSM Weyl fermions. After EWSB, one can get the following $4\times 4$ neutralino mass matrix and the diagonalization procedure
\begin{eqnarray}
M_{\chi^0} &=& \left(
\begin{array}{cccc}
M^{B}_D &0 &-\frac{1}{2} g_1 v_d  &\frac{1}{2} g_1 v_u \\
0 &M^{W}_D &\frac{1}{2} g_2 v_d  &-\frac{1}{2} g_2 v_u \\
- \frac{1}{\sqrt{2}} \lambda_d v_d  &-\frac{1}{2} \Lambda_d v_d  &-\mu_d^{eff,+}&0\\
\frac{1}{\sqrt{2}} \lambda_u v_u  &-\frac{1}{2} \Lambda_u v_u  &0 &\mu_u^{eff,-}\end{array}
\right),(N^{1})^{\ast} M_{\chi^0} (N^{2})^{\dagger} = M_{\chi^0}^{diag},
 \end{eqnarray}
where $N^1$ and $N^2$ are 4$\times$4 unitary matrices. The $\mu_i$ are given by,
\begin{equation}
\begin{split}
\mu_d^{eff,+}&= \frac{1}{2} \Lambda_d v_T  + \frac{1}{\sqrt{2}} \lambda_d v_S  + \mu_d ,\\
\mu_u^{eff,-}&= -\frac{1}{2} \Lambda_u v_T  + \frac{1}{\sqrt{2}} \lambda_u v_S  + \mu_u.
\end{split}
\end{equation}
The $v_T$ and $v_S$ are vacuum expectation values of $\hat{T}$ and $\hat{S}$ which carry zero $R$-charge.The chargino mass matrix and the diagonalization procedure is given by,
\begin{equation}
M_{\chi^{\pm}} = \left(
\begin{array}{cc}
g_2 v_T  + M^{W}_D &\frac{1}{\sqrt{2}} \Lambda_d v_d \\
\frac{1}{\sqrt{2}} g_2 v_d  &-\frac{1}{2} \Lambda_d v_T  + \frac{1}{\sqrt{2}} \lambda_d v_S  + \mu_d\end{array}
\right),(U^{1})^{\ast} M_{\chi^{\pm}} (V^{1})^{\dagger} =M_{\chi^{\pm}}^{diag},
 \end{equation}
where $U^1$ and $V^1$ are 2$\times$2 unitary matrices.

In the MRSSM, LFV decays mainly originate from the potential misalignment in slepton mass matrices. In the gauge eigenstate basis $\tilde{\nu}_{iL}$, the sneutrino mass matrix and the diagonalization procedure are
\begin{equation}
M^2_{\tilde{\nu}} =
m_{\tilde{l}}^2+\frac{1}{8}(g_1^2+g_2^2)( v_{d}^{2}- v_{u}^{2})+g_2 v_T M^{W}_D-g_1 v_S M^{B}_D,
Z^V M^2_{\tilde{\nu}} (Z^{V})^{\dagger} = M^{2,\textup{diag}}_{\tilde{\nu}},\label{sn}
\end{equation}
where the last two terms in mass matrix are newly introduced by the MRSSM. The slepton mass matrix and the diagonalization procedure are
\begin{equation}
M^2_{\tilde{e}} = \left(
\begin{array}{cc}
(M^2_{\tilde{e}})_{LL} &0 \\
0  &(M^2_{\tilde{e}})_{RR}\end{array}
\right),Z^E M^2_{\tilde{e}} (Z^{E})^{\dagger} =M^{2,\textup{diag}}_{\tilde{e}},\label{sl}
 \end{equation}
where
\begin{equation}
\begin{split}
(M^2_{\tilde{e}})_{LL} &=m_{\tilde{l}}^2+ \frac{1}{2} v_{d}^{2} |Y_{e}|^2 +\frac{1}{8}(g_1^2-g_2^2)(v_{d}^{2}- v_{u}^{2}) -g_1 v_S M_D^B-g_2v_TM_D^W ,\\
(M^2_{\tilde{e}})_{RR} &= m_{\tilde{r}}^2+\frac{1}{2}v_d^2|Y_e|^2+\frac{1}{4}g_1^2( v_{u}^{2}- v_{d}^{2})+2g_1v_SM_D^B.
\end{split}
\end{equation}
The sources of LFV are the off-diagonal entries of the $3\times 3$ soft supersymmetry breaking matrices $m_l^2$ and $m_r^2$ in Eqs.(\ref{sn}, \ref{sl}). From Eq.(\ref{sl}) we can see that the left-right slepton mass mixing is absent in the MRSSM, whereas the $A$ terms are present in the MSSM.

The mass matrix for up squarks and down squarks, and the relevant diagonalization procedure are
\begin{equation}
\begin{split}
M^2_{\tilde{u}} &= \left(
\begin{array}{cc}
(M^2_{\tilde{u}})_{LL} &0 \\
0  &(M^2_{\tilde{u}})_{RR}
\end{array}
\right), Z^U M^2_{\tilde{u}} (Z^{U})^{\dagger} =M^{2,\textup{diag}}_{\tilde{u}}, \\
M^2_{\tilde{d}} &= \left(
\begin{array}{cc}
(M^2_{\tilde{d}})_{LL} &0 \\
0  &(M^2_{\tilde{d}})_{RR}
\end{array}
\right),Z^D M^2_{\tilde{d}} (Z^{D})^{\dagger} =M^{2,\textup{diag}}_{\tilde{d}},\label{sud}
\end{split}
\end{equation}
where
\begin{equation}
\begin{split}
(M^2_{\tilde{u}})_{LL} &=m_{\tilde{q}}^2+ \frac{1}{2} v_{u}^{2} |Y_{u}|^2
+\frac{1}{24}(g_1^2-3g_2^2)(v_{u}^{2}- v_{d}^{2}) +\frac{1}{3}g_1 v_S M_D^B+g_2v_TM_D^W ,\\
(M^2_{\tilde{u}})_{RR} &= m_{\tilde{u}}^2+\frac{1}{2}v_u^2|Y_u|^2+\frac{1}{6}g_1^2( v_{d}^{2}- v_{u}^{2})-\frac{4}{3} g_1v_SM_D^B,\\
(M^2_{\tilde{d}})_{LL} &=m_{\tilde{q}}^2+ \frac{1}{2} v_{d}^{2} |Y_{d}|^2
+\frac{1}{24}(g_1^2+3g_2^2)(v_{u}^{2}- v_{d}^{2}) +\frac{1}{3}g_1 v_S M_D^B-g_2v_TM_D^W ,\\
(M^2_{\tilde{d}})_{RR} &= m_{\tilde{d}}^2+\frac{1}{2}v_d^2|Y_d|^2+\frac{1}{12}g_1^2( v_{u}^{2}- v_{d}^{2})+\frac{2}{3} g_1v_SM_D^B.
\end{split}
\end{equation}

The explicit expressions of the Feynman rules between fermions, sfermions and neutralinos/charginos are given as
\begin{equation}
\begin{split}
-i{\cal L}= &\bar{l}_i[Y_e^iZ^{V\ast}_{ki}U^{1\ast}_{j2}P_L-g_2V^{1}_{j1}Z^{V\ast}_{ki}P_R]\chi^{\pm}\tilde{\nu}_{k}\\
& -\bar{l}_i[\sqrt{2}g_1Z^{E\ast}_{k(3+i)}N^{1\ast}_{j1}P_L+Y_e^{i}N^{2}_{j3}Z^{E\ast}_{k(3+i)}P_R]\chi^0_j\tilde{e}_k\\
& -\bar{l}_i[Y_e^{i}N^{2\ast}_{j3}Z^{E\ast}_{ki}P_L-\frac{1}{\sqrt{2}}Z^{E\ast}_{ki}(g_1N^{1}_{j1}+g_2N^{1}_{j2})P_R]\chi^{0c}_j\tilde{e}_k\\
&-\bar{d}_i[\frac{2\sqrt{2}}{3}g_1N^{1\ast}_{j1}Z^{D\ast}_{k(3+i)}P_L+Y^i_dZ^{D\ast}_{k(3+i)}N^2_{j3}P_R]\chi^0_j\tilde{d}_k\\
&-\bar{d}_i[Y^i_dN^{2\ast}_{j3}Z^{D\ast}_{ki}P_L+\frac{\sqrt{2}}{6}Z^{D\ast}_{ki}(g_1N^1_{j1}-3g_2N^1_{j2})P_R]\chi^{0c}_j\tilde{d}_k\\
&+\bar{d}_i[Y^i_dU^{1\ast}_{j2}Z^{U\ast}_{ki}P_L-g_2Z^{U\ast}_{ki}V^1_{j1}P_R]\chi^{\pm}_j\tilde{u}_k,
\end{split}\label{rule1}
\end{equation}
where all the repeated indices of generation should be summed over.

\begin{figure*}
\centering
\includegraphics[width=6.5in]{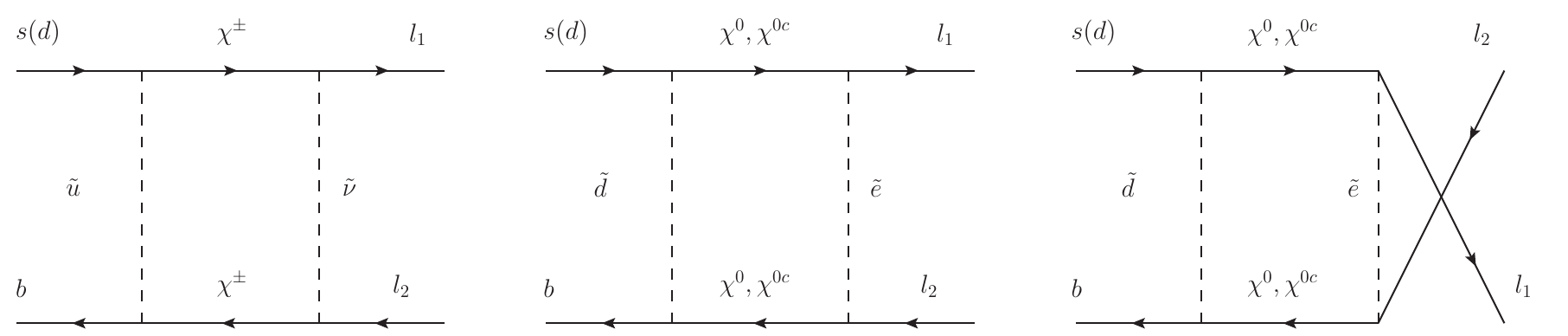}
\caption[]{The Feynman diagrams contributing to $B^0_q\rightarrow l_1l_2$ in the MRSSM.}
\label{Diag}
\end{figure*}

In the MRSSM, the LFV decays $B^0_q\rightarrow l_1l_2$ arises at the box level as shown in FIG.\ref{Diag}.
With the effective lagrangian method, these processes are described by the four-fermion interaction lagrangian as
\begin{eqnarray}
{\cal L}&=&\sum_{\alpha=S,V,T;\beta,\delta=L,R} B^{\alpha}_{\beta\delta}\bar{l}_1\Gamma_\alpha P_{\beta}l_2\bar{d}_K\Gamma_\alpha P_{\delta}d_L+h.c.,
\label{eff}
\end{eqnarray}
where the subscripts $K$, $L$ denote the quark flavors. The symbol $\Gamma_\alpha$ denotes the Dirac matrix structure and $\Gamma_S$ = 1, $\Gamma_V$ = $\gamma_\mu$ and $\Gamma_T$ = $\sigma_{\mu\nu}$. The symbols $P_{L/R}$ denote the chirality projectors. The amplitude $\mathcal{M}$ in FIG.\ref{Diag} is composed of several form factors
\begin{eqnarray}
(4\pi)^2\mathcal{M}=F_S\bar{l}_1l_2+F_P\bar{l}_1\gamma^5l_2+F_V p_{\mu}\bar{l}_1\gamma^\mu l_2+F_Ap_\mu\bar{l}_1\gamma^\mu\gamma^5 l_2,
\label{M}
\end{eqnarray}
where the form factors $F_S$, $F_P$, $F_V$ and $F_A$ are combinations of the Wilson coefficients $B^{\alpha}_{\beta\delta}$,
\begin{equation}
\begin{split}
F_S&=\frac{i m^2_{B^0_q}f_{B^0_q}}{4(m_b+m_q)}(B^S_{LL}+B^S_{LR}-B^S_{RR}-B^S_{RL}),\\
F_P&=\frac{i m^2_{B^0_q}f_{B^0_q}}{4(m_b+m_q)}(-B^S_{LL}+B^S_{LR}-B^S_{RR}+B^S_{RL}),\\
F_V&=-\frac{i f_{B^0_q}}{4}(B^V_{LL}+B^V_{LR}-B^V_{RR}-B^V_{RL}),\\
F_A&=-\frac{i f_{B^0_q}}{4}(-B^V_{LL}+B^V_{LR}-B^V_{RR}+B^V_{RL}),
\end{split}
\end{equation}
where $m_{b}$, $m_{q}$ and $m_{B^0_q}$ denote the mass of $b$ quark, $d/s$ quark and $B^0_{d/s}$, respectively, and $f_{B^0_q}$ is the decay constant of $B^0_{d/s}$.
The coefficients in FIG.\ref{Diag} are calculated to be
\begin{equation}
\begin{split}
B^S_{LL}&=\frac{1}{32\pi^2}C^{qS_1F_1}_LC^{bS_1F_2*}_LC^{l_1S_2F_1}_LC^{l_2S_2F_2*}_L m_{F_1} m_{F_2}D_0( m_{F_2}^2, m_{F_1}^2, m_{S_1}^2, m_{S_2}^2),\\
B^S_{LR}&=\frac{1}{8\pi^2}C^{qS_1F_1}_LC^{bS_1F_2*}_LC^{l_1S_2F_1}_RC^{l_2S_2F_2*}_RD_{00}(m_{F_2}^2, m_{F_1}^2, m_{S_1}^2, m_{S_2}^2),\\
B^V_{LL}&=-\frac{1}{16\pi^2}C^{qS_1F_1}_LC^{bS_1F_2*}_RC^{l_1S_2F_1}_RC^{l_2S_2F_2*}_L m_{F_1} m_{F_2}D_0(m_{F_2}^2, m_{F_1}^2, m_{S_1}^2, m_{S_2}^2),\\
B^V_{LR}&=\frac{1}{32\pi^2}C^{qS_1F_1}_LC^{bS_1F_2*}_RC^{l_1S_2F_1}_LC^{l_2S_2F_2*}_RD_{00}(m_{F_2}^2, m_{F_1}^2, m_{S_1}^2, m_{S_2}^2),
\end{split}\label{D0D00}
\end{equation}
where $F_1F_2S_1S_2\in\{$$\chi^{\pm}\chi^{\pm}\tilde{u}\tilde{\nu}$, $\chi^{0}\chi^{0}\tilde{d}\tilde{e}$, $\chi^{0c}\chi^{0}\tilde{d}\tilde{e}$, $\chi^{0}\chi^{0c}\tilde{d}\tilde{e}$, $\chi^{0c}\chi^{0c}\tilde{d}\tilde{e}\}$. The coefficients $C^{\{qS_1F_1, bS_1F_2, ...\}}$ denote the interaction between quark/lepton ($q$, $b$, $l_1$, $l_2$), scalar particle ($S_1$, $S_2$) and fermion ($F_1$, $F_2$), respectively. The Wilson coefficients are left-right symmetric, i.e, $B^S_{RR}=B^S_{LL}(L\leftrightarrow R)$, $B^S_{RL}=B^S_{LR}(L\leftrightarrow R)$, $B^V_{RR}=B^V_{LL}(L\leftrightarrow R)$ and $B^V_{RL}=B^V_{LR}(L\leftrightarrow R)$. The explicit expressions of the loop integrals $D_0$ and $D_{00}$ in Eq(\ref{D0D00}) are given as \cite{Dreiner}
\begin{equation}
\begin{split}
D_ {0}  (x,y,z,t)&=-[\frac {y \log\frac{y}{x}}{(y-x)(y-z)(y-t)} +\frac {z \log\frac{z}{x}}{(z-x)(z-y)(z-t)}+\frac {t \log\frac{t}{x}}{(t-x)(t-y)(t-z)}],\\
D_ {00}  (x,y,z,t)&=- \frac {1}{4}[\frac {y^ 2 \log\frac{y}{x}}{(y-x)(y-z)(y-t)} +\frac {z^ 2 \log\frac{z}{x}}{(z-x)(z-y)(z-t)}+\frac {t^ 2 \log\frac{t}{x}}{(t-x)(t-y)(t-z)}].
\end{split}
\end{equation}

From Eq.(\ref{M}) one can easily calculate the squared amplitude
\begin{equation}
\begin{split}
|\mathcal{M}|^2=&\frac{1}{128\pi^4}\big(|F_S|^2(m^2_{B^0_q}-(m_2+m_1)^2)+|F_P|^2(m^2_{B^0_q}-(m_1-m_2)^2)\\
&+|F_V|^2(m^2_{B^0_q}(m_2-m_1)^2-(m_2-m_1)^2)+|F_A|^2(m^2_{B^0_q}(m_2+m_1)^2\\
&-(m_2-m_1)^2)+2 Re(F_S F_V^*)(m_1-m_2)(m^2_{B^0_q}+(m_2+m_1)^2)\\
&+2 Re(F_P F_A^*)(m_1+m_2)(m^2_{B^0_q}-(m_2-m_1)^2)\big).
\end{split}
\end{equation}
The analytic expression of the branching ratio of $B^0_q\rightarrow\bar{l}_1l_2$ is given by
\begin{eqnarray}
\mathrm{BR}(B^0_q\rightarrow\bar{l}_1l_2)=\frac{\tau_{B^0_q}}{16\pi M_{B^0_q}}\sqrt{1-(\frac{m_2+m_1}{M^2_{B^0_q}})^2}\sqrt{1-(\frac{m_2-m_1}{M^2_{B^0_q}})^2}
|\mathcal{M}|^2,
\end{eqnarray}
where $\tau_{B^0_q}$ is the life time of $B^0_{d/s}$. The total BR($B^0_q\rightarrow l_1l_2$) is the sum BR($B^0_q\rightarrow\bar{l}_1l_2$) + BR($B^0_q\rightarrow l_1\bar{l}_2$).

\section{Numerical Analysis\label{sec3}}

The numerical calculations of BR($B^0_q\rightarrow l_1l_2$) in the MRSSM are performed using BSMArts \cite{Goodsell} and the SARAH family of tools \cite{SARAH, SARAH1, SARAH2, SPheno1, SPheno2, Flavor, Flavor2,Bahl,Feroz1,Feroz2}. In the numerical analysis, we adopt the following values for the parameters of meson $B^0_q$ \cite{PDG}
\begin{equation}
\begin{split}
m_{B^0_s}=5.36691 \mathrm{GeV}, f_{B^0_s}=227 \mathrm{MeV}, \tau_{B^0_s}=1.516\times 10^{-12} \mathrm{s},\\
m_{B^0_d}=5.27963 \mathrm{GeV}, f_{B^0_d}=190 \mathrm{MeV}, \tau_{B^0_d}=1.517\times 10^{-12} \mathrm{s}.
\end{split}
\end{equation}
To decrease the number of free parameters involved in our calculation, we adopt the following values for the model parameters
\begin{equation}
\begin{split}
&M_D^O=m_O=1500,m_{R_d}=m_{R_u}=2000,\\
&(m^2_{\tilde{l}})_{ii}=(m^2_{\tilde{r}})_{ii}=1000^2,(i=1,2,3),\\
&(m^2_{\tilde{q}})_{ii}=(m^2_{\tilde{u}})_{ii}=(m^2_{\tilde{d}})_{ii}=2500^2,(i=1,2),\\
&(m^2_{\tilde{q}})_{33}=(m^2_{\tilde{u}})_{33}=(m^2_{\tilde{d}})_{33}=1000^2, m_T=3000,
\end{split}\label{N1}
\end{equation}
which are taken from Refs. \cite{Die3, Die5}. It is worth mentioning that the off-diagonal elements of the squark mass matrices $m^2_{\tilde{q}}$, $m^2_{\tilde{u}}$, $m^2_{\tilde{d}}$, and the slepton mass matrices $m^2_{\tilde{l}}$, $m^2_{\tilde{r}}$ in Eq.(\ref{N1}) are assumed to be zero, which implies the absence of flavor mixing in the squark and slepton sectors. In this paper, these off-diagonal entries are parameterized by mass insertion as in \cite{MIn1,MIn2}
\begin{equation}
\Big(m^{2}_{X}\Big)_{ij}=\delta ^{ij}_{X}\sqrt{(m^{2}_{X})_{ii}(m^{2}_{X})_{jj}},\;\;i\neq j,
\end{equation}
where X $\in$ ($\tilde{q}$, $\tilde{u}$, $\tilde{d}$, $\tilde{l}$, $\tilde{r}$) and $i,j=1,2,3$. For simplicity, we assume that $\delta ^{ij}_{\tilde{l}}$ = $\delta ^{ij}_{\tilde{r}}$ $\equiv$ $\delta ^{ij}_L$ and $\delta ^{ij}_{\tilde{q}}$ = $\delta ^{ij}_{\tilde{u}}$ = $\delta ^{ij}_{\tilde{d}}$ $\equiv$ $\delta ^{ij}_Q$ . 

Before scanning over the parameter space, several constraints are applied. To ensure the model can accommodate a Higgs boson with mass $m_{h}^{exp}$ around 125 GeV, the lightest Higgs boson with mass $m_{h}^{th}$ in the MRSSM is chosen to be similar to the one in the SM. The predicted mass of the W-boson in the MRSSM and the low-energy observables related to B meson physics require to be consistent with experimental measurements \cite{PDG}. Constraints from electroweak precision observables require that the oblique parameters S, T, and U \cite{Peskin1,Peskin2} to matching the global fit results as detailed in \cite{PDG}. The mentioned constraints are all listed in the TABLE.\ref{constraints} .
\begin{table}[h]
\caption{Summary of constraints considered in this paper. }
\begin{tabular}{@{}c|c|c|c@{}} \toprule
Constraint&Range&Constraint&Range\\
\colrule
$m_W$& 80.3692 $\pm$ 0.0133&$m_{h}^{exp}$& 125.2 $\pm$ 0.11\\
S&-0.04 $\pm$ 0.10&T& 0.01 $\pm$ 0.12\\
U&-0.01 $\pm$ 0.09&BR($B\rightarrow X_s \gamma$)&$(3.49\pm 0.19)\times 10^{-4}$ \\
BR($B^0_s\rightarrow \mu\mu$)&$(3.34\pm 0.27) \times 10^{-9}$&BR($B^0_d\rightarrow \mu\mu$)&<$1.5\times 10^{-10}$\\ \botrule
\end{tabular}
\label{constraints}
\end{table}

\begin{table}[h]
\caption{Ranges of input parameters. }
\begin{tabular}{@{}c|c|c|c|c|c@{}} \toprule
Parameter&Prior&Range&Parameter&Prior&Range\\
\colrule
$\tan\beta$&flat&3 $\sim$ 50&B$_\mu$&flat&$100^2 \sim 600^2$\\
$\lambda_d$&flat&-2 $\sim$ 2&$\lambda_u$&flat&-2 $\sim$ 2\\
$\Lambda_d$&flat&-2 $\sim$ 2&$\Lambda_u$&flat&-2 $\sim$ 2\\
$M_D^B$&flat&500 $\sim$ 700&$M_D^W$&flat&500 $\sim$ 700\\
$\mu_d$&flat&400 $\sim$ 600&$\mu_u$&flat&400 $\sim$ 600\\
$m_S$&flat&1000 $\sim$ 3000&$\delta^{12}_{Q}$&flat&0.01 $\sim$ 1\\
$\delta^{13}_{Q}$&flat&0.01 $\sim$ 1&$\delta^{23}_{Q}$&flat&0.01 $\sim$ 1\\ \botrule
\end{tabular}
\label{scan}
\end{table}

\begin{figure*}[htbp]
\centering
\includegraphics[width=7.0in]{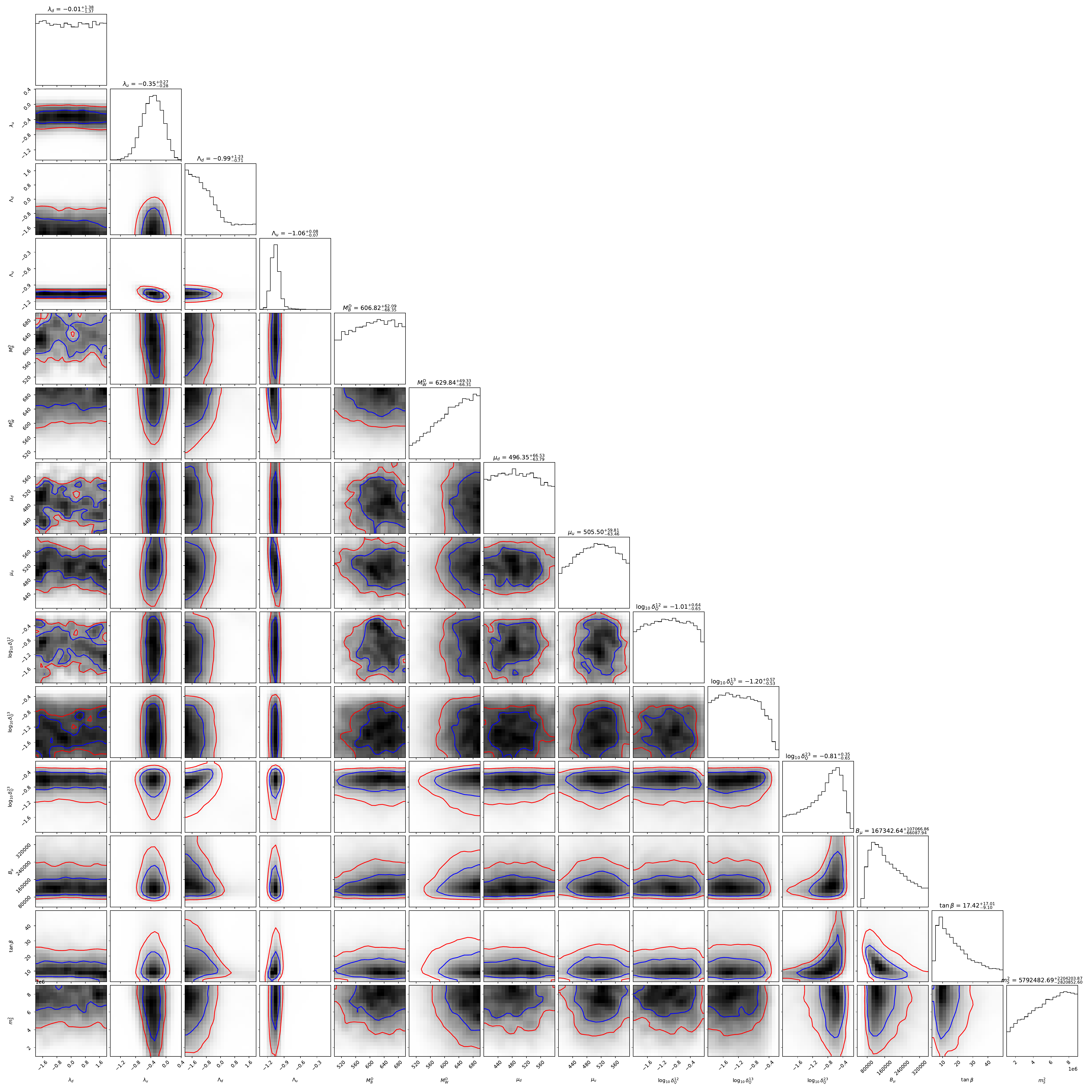}
\caption[]{Corner plot showing the sensitivity of the SM-like higgs mass to the 14 parameters in TABLE.\ref{scan} in the MRSSM.}
\label{cornerplot}
\end{figure*}

We perform scans over the 14 parameters, as shown in TABLE.\ref{scan}, using flat priors for all of them. The fit of $m_{h}^{th}$ with $m_{h}^{exp}$ is performed with BSMArts \cite{Goodsell} and make use of the HiggsTools \cite{Bahl} and MultiNest \cite{Feroz1,Feroz2}. The p-value reported by HiggsTools is chosen to be larger than 5\%.  In FIG.\ref{cornerplot}, we show the results of MultiNest analysis in the form of corner plots by use of Corner \cite{Mackey}. Regions enclosed within the blue and red lines can explain the SM-like higgs mass at 1$\sigma$ and 1.5$\sigma$ levels, respectively. The 14 parameters, which can explain the SM-like higgs mass at 1$\sigma$ level, are given by 
\begin{equation}
\begin{split}
&B_\mu=167342.64^{+107066.86}_{-66087.94},\tan\beta=17.42^{+17.01}_{-9.10},M_S^2=5792482.69^{+2204203.87}_{-2820852.60},\\
&\lambda_d=-0.01^{+1.38}_{-1.37},\lambda_u=-0.35^{+0.27}_{-0.28},\Lambda_d=-0.99^{+1.23}_{-0.71},\Lambda_u=-1.06^{+0.08}_{-0.07},\\
&M_B^D=606.82^{+62.09}_{-66.31},M_W^D=629.84^{+49.33}_{-66.31},
\mu_d=496.35^{+66.53}_{-63.79},\mu_u=505.50^{+59.81}_{-63.46},\\
&log_{10}\,\delta_Q^{12}=-1.01^{+0.64}_{-0.65},log_{10}\,\delta_Q^{13}=-1.2^{+0.57}_{-0.53},log_{10}\,\delta_Q^{23}=-0.81^{+0.35}_{-0.65}.
\end{split}\label{space}
\end{equation}

\begin{figure}[htbp]
\setlength{\unitlength}{1mm}
\centering
\begin{minipage}[c]{1\textwidth}
\includegraphics[width=5.5in]{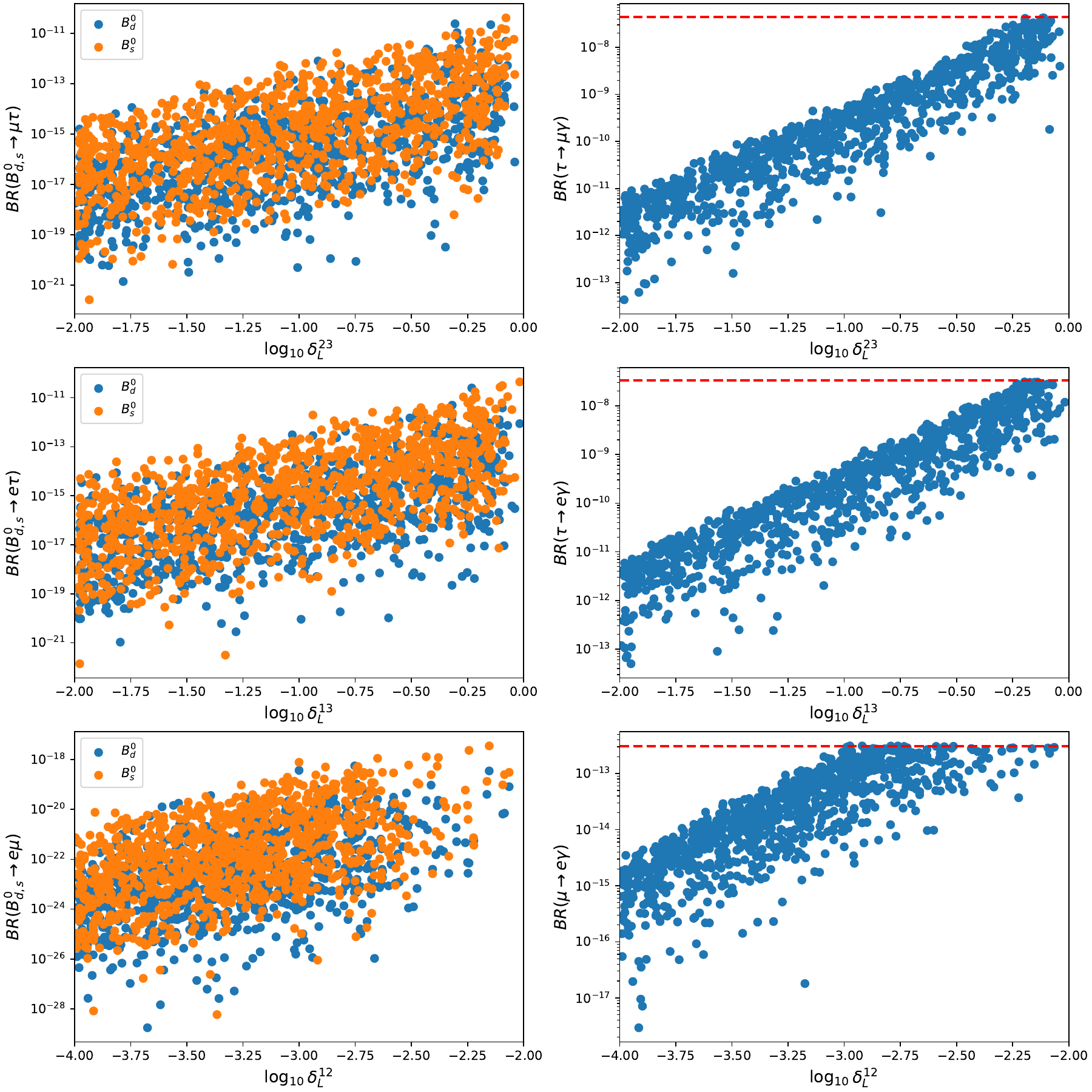}
\end{minipage}
\caption[]{Scatter plot showing the dependence of BR($B^0_q\rightarrow l_1l_2$) and BR($l_2\rightarrow l_1\gamma$) on $\delta^{ij}_L$ in the MRSSM. }
\label{lfv}
\end{figure}

\begin{table}[h]
\caption{Ranges of $\delta ^{ij}_L$. }
\begin{tabular}{@{}c|c|c|c|c|c|c|c|c@{}} \toprule
Parameter&Prior&Range&Parameter&Prior&Range&Parameter&Prior&Range\\
\colrule
$\delta^{12}_{L}$&flat&$10^{-4}$ $\sim$ 1&$\delta^{13}_{L}$&flat&0.01 $\sim$ 1&$\delta^{23}_{L}$&flat&0.01 $\sim$ 1\\ \botrule
\end{tabular}
\label{scanL}
\end{table}
\begin{table}[h]
\caption{Current limits on BR($l_2\rightarrow l_1\gamma$). }
\begin{tabular}{@{}c|c|c|c|c|c@{}} \toprule
Decay&Limit&Decay&Limit&Decay&Limit\\
\colrule
$\mu\rightarrow e\gamma$&$3.1\times 10^{-13}$&$\tau\rightarrow e\gamma$&$3.3\times 10^{-8}$&$\tau\rightarrow \mu\gamma$&$4.2\times 10^{-8}$\\\botrule
\end{tabular}
\label{constraints1}
\end{table}

\begin{figure}[htbp]
\setlength{\unitlength}{1mm}
\centering
\begin{minipage}[c]{1\textwidth}
\includegraphics[width=6.50in]{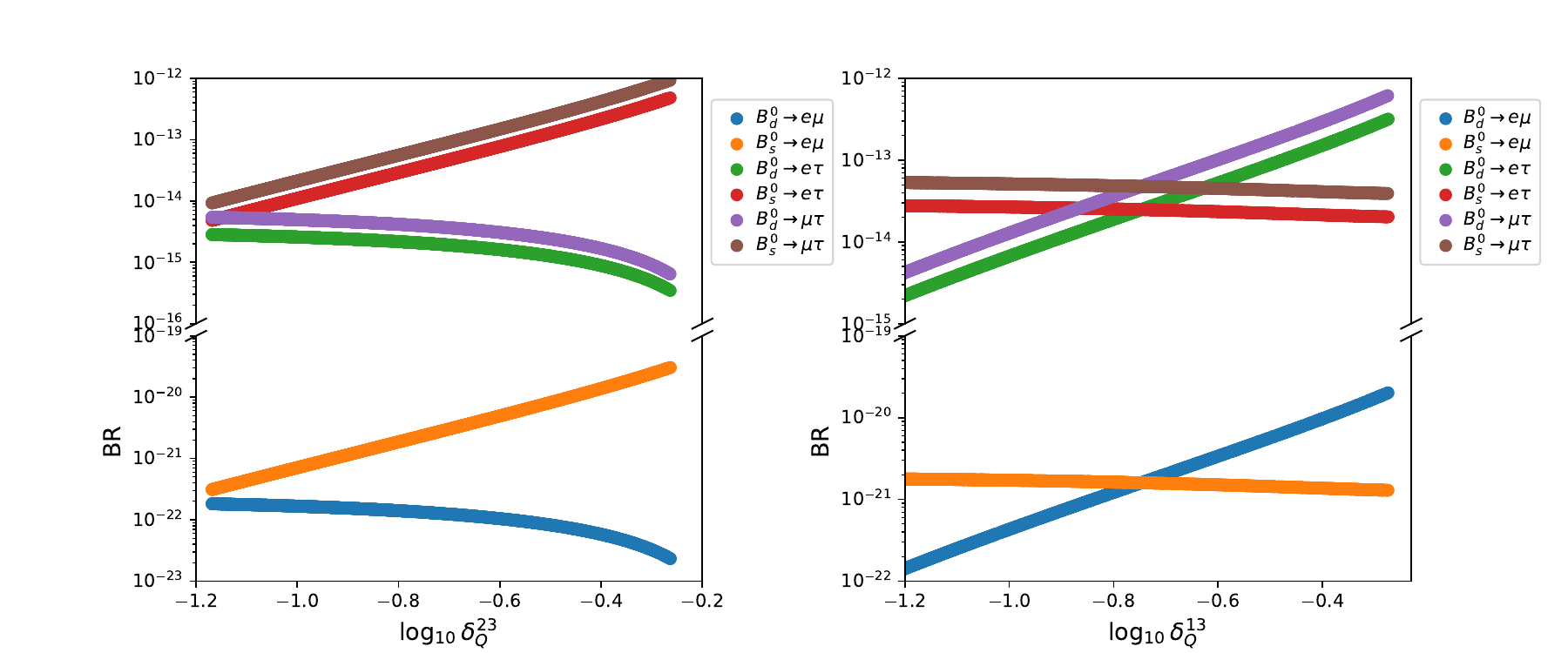}
\end{minipage}
\caption[]{Plot showing the dependence of BR($B^0_q\rightarrow l_1l_2$) on $\delta^{ij}_Q$ in the MRSSM.}
\label{qfv}
\end{figure}

We perform scans over the 17 parameters in TABLE.\ref{scan} and TABLE.\ref{scanL}, and plot the predictions of BR($B^0_q\rightarrow l_1l_2$) versus log$_{10}\,\delta^{ij}_L$ in FIG.\ref{lfv}, where the corresponding predictions for BR($l_2\rightarrow l_1\gamma$) are also presented.  All points satisfy the constraints in TABLE.\ref{constraints} and the current experimental limits in TABLE.\ref{constraints1} \cite{PDG}. The red horizontal lines denote the current experimental bounds of BR($l_2\rightarrow l_1\gamma$). In each subfigure, only the indicated $\delta^{ij}_L$ is varied with all other $\delta^{ij}_L$ set to zero since they have no effect on the prediction. Thus, there are 15 parameters for each scan. Similar to the results in Refs.\cite{susy3,ss1,ss3,ss4,ss5}, the predictions for BR($B^0_q\rightarrow e\mu$), BR($B^0_q\rightarrow e\tau$), and BR($B^0_q\rightarrow \mu\tau$) are affected by the mass insertions $\delta^{12}_L$, $\delta^{13}_L$, and $\delta^{23}_L$, respectively. The prediction of BR($B^0_q\rightarrow e\tau$) and BR($B^0_q\rightarrow \mu\tau$) goes up to $\mathcal{O}(10^{-10})$. This value is five orders of magnitude below the current experimental limit and four orders of magnitude below the expected future limit \cite{Altmannshofer}. The prediction of BR($B^0_q\rightarrow e\mu$) is eight orders of magnitude below the future experimental limit \cite{Aaij}. In the following discussion, the default values log$_{10}\,\delta^{12}_L=-2.6$ for BR($B^0_q\rightarrow e\mu$), log$_{10}\,\delta^{13}_L= - 0.2$ for BR($B^0_q\rightarrow e\tau$) and log$_{10}\,\delta^{23}_L= - 0.1$ for BR($B^0_q\rightarrow \mu\tau$) are used by default.

Taking the central values in Eq.(\ref{space}) as default, we plot the predictions of BR($B^0_q\rightarrow l_1l_2$) versus log$_{10}\,\delta^{13}_Q$, BR($B^0_q\rightarrow l_1l_2$) versus log$_{10}\,\delta^{23}_Q$ in FIG.\ref{qfv}. In each subfigure, only the indicated $\delta^{ij}_Q$ is varied with all other $\delta^{ij}_Q$ set to the central values in Eq.(\ref{space}). Both $\delta^{23}_Q$ and $\delta^{13}_Q$  have significant impacts on BR($B^0_q\rightarrow l_1l_2$). The predicted BR($B^0_d\rightarrow l_1l_2$) increase as the parameter $\delta^{13}_Q$ increases and decrease as the parameter $\delta^{23}_Q$ increases, whereas the predicted BR($B^0_s\rightarrow l_1l_2$) behave exactly opposite to BR($B^0_d\rightarrow l_1l_2$). The parameters $\delta^{ij}_Q$ may play different roles in the LFV decay of mesons. The effect from $\delta^{ij}_Q$ would be too small to be neglected for mesons those containing two same generation quarks, e.g., $\phi$, $J/\Psi$, and $\Upsilon (nS)$\cite{ss3}. For mesons those containing two different generation quarks, the effect from $\delta^{ij}_Q$ may not be neglected, e.g., $B^0_d$ and $B^0_s$\cite{susy3}.

\begin{figure}[htbp]
\setlength{\unitlength}{1mm}
\centering
\begin{minipage}[c]{1\textwidth}
\includegraphics[width=4.0in]{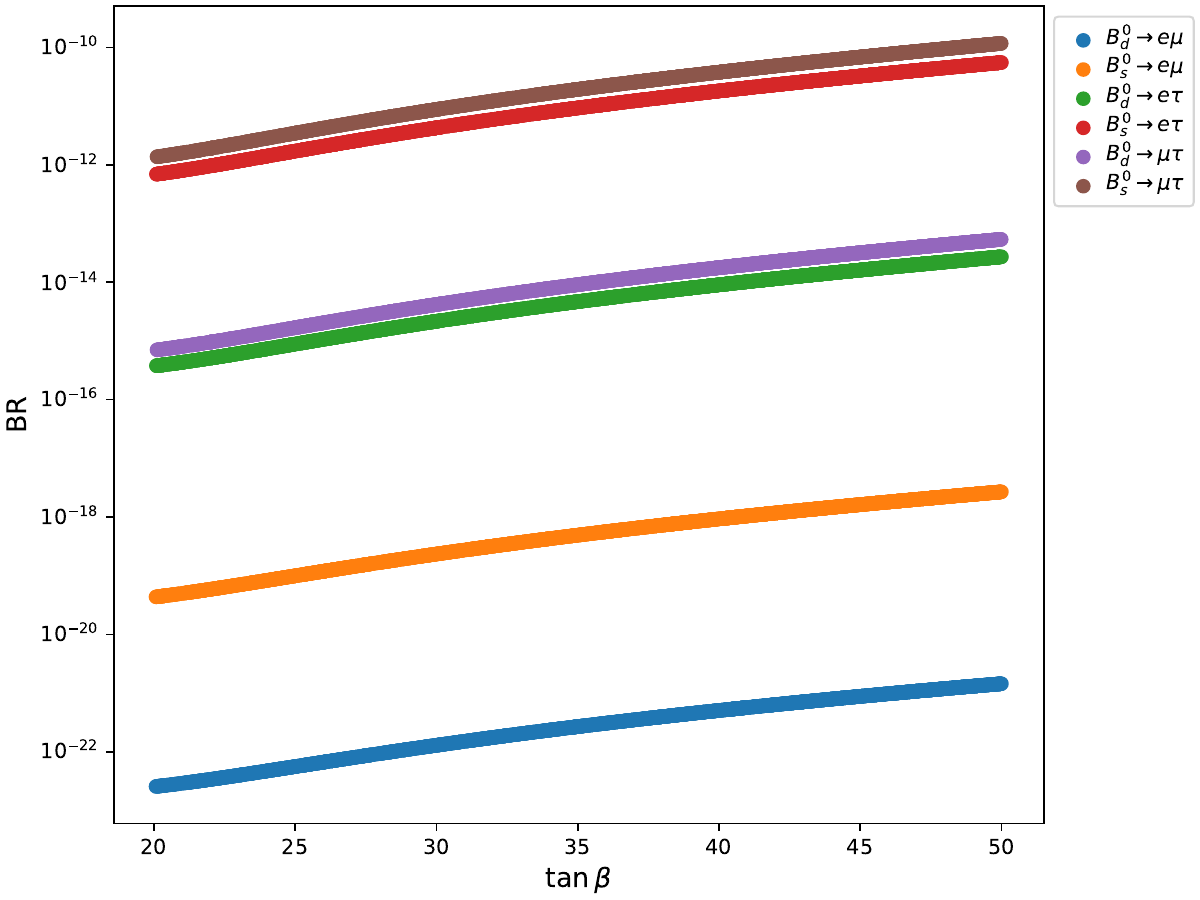}
\end{minipage}
\caption[]{Plot showing the dependence of BR($B^0_q\rightarrow l_1l_2$) on $\tan\beta$ in the MRSSM.}
\label{tanbeta}
\end{figure}
Taking the central values in Eq.(\ref{space}) and log$_{10}\,\delta^{23}_Q=-0.25$ as default, we plot the predictions of BR($B^0_q\rightarrow l_1l_2$) versus $\tan \beta$ in FIG.\ref{tanbeta}. It shows that $\tan \beta$ has an significant impact on BR($B^0_q\rightarrow l_1l_2$) as well and the predicted BR($B^0_d\rightarrow l_1l_2$) increase as $\tan \beta$ increases. At $\tan \beta\sim 50$, the predicted BR($B^0_s\rightarrow e\tau$) and BR($B^0_s\rightarrow \mu\tau$) might be enhanced up to around $\mathcal{O}(10^{-10})$. It is noted that the default value of $\tan \beta$ is 17.42 in FIG.\ref{qfv}. Thus, the upper prediction for BR($B^0_q\rightarrow l_1\tau$) is around $\mathcal{O}(10^{-12})$ in FIG.\ref{qfv}.

\begin{figure}[htbp]
\setlength{\unitlength}{1mm}
\centering
\begin{minipage}[c]{1\textwidth}
\includegraphics[width=5.5in]{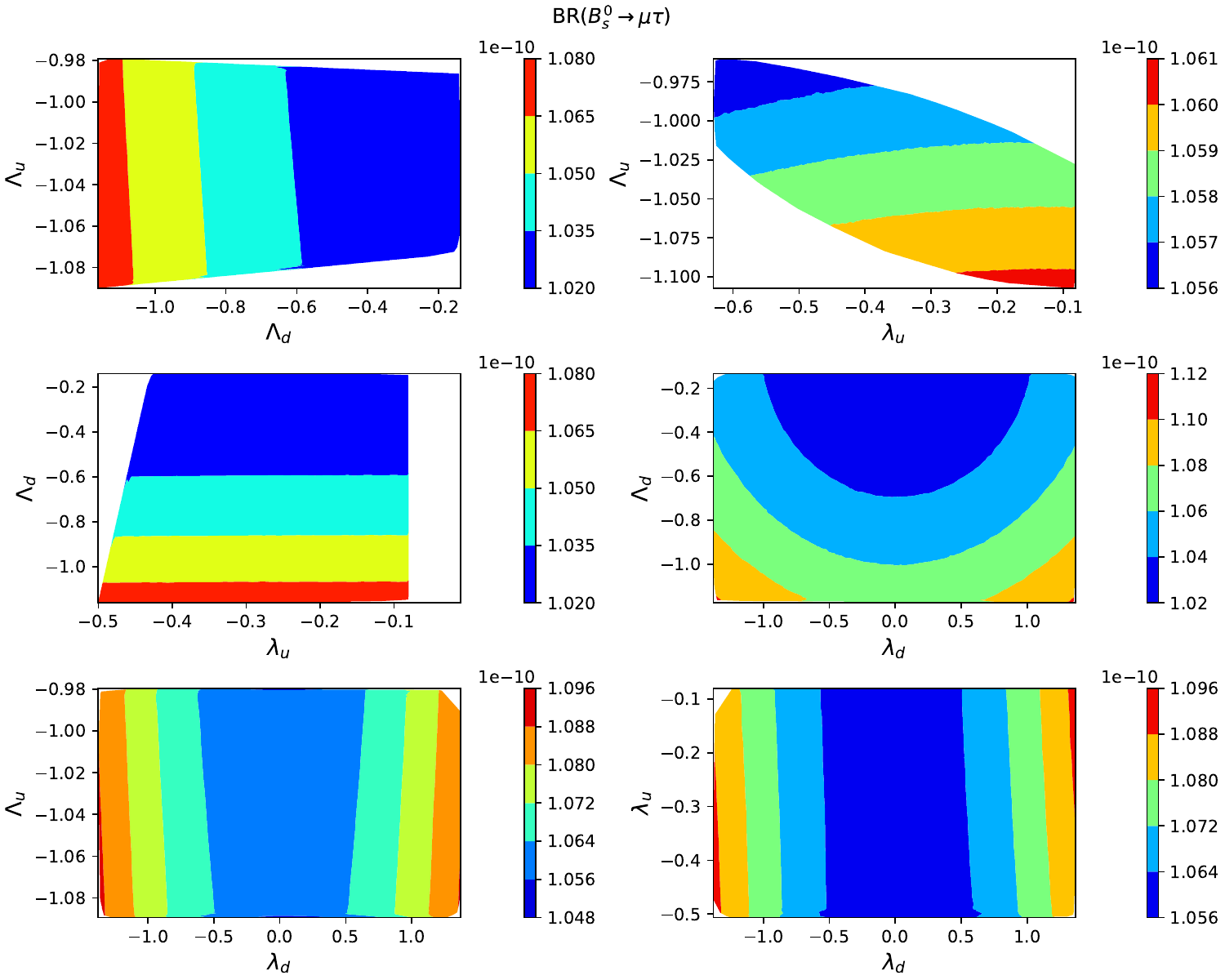}
\end{minipage}
\caption[]{Contour plot showing the dependence of BR($B^0_s\rightarrow \mu\tau$) on $\lambda_u$, $\lambda_d$, $\Lambda_u$ and $\Lambda_d$ in the MRSSM. }
\label{LSTUD}
\end{figure}
\begin{figure}[htbp]
\setlength{\unitlength}{1mm}
\centering
\begin{minipage}[c]{1\textwidth}
\includegraphics[width=5.5in]{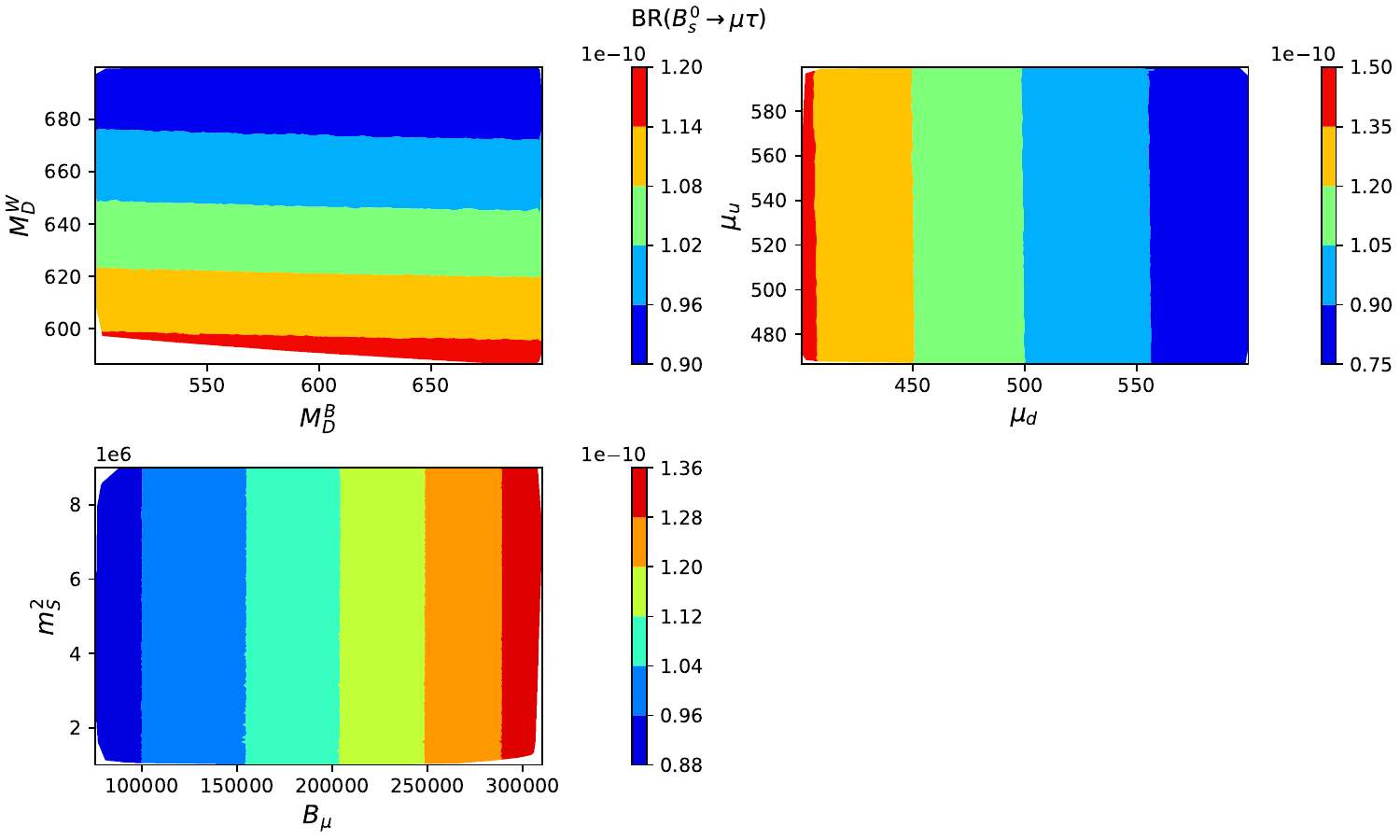}
\end{minipage}
\caption[]{Contour plot showing the dependence of BR($B^0_s\rightarrow \mu\tau$) on $M_D^W$, $M_D^B$, $\mu_u$, $\mu_d$, $B_\mu$ and $m_S$ in the MRSSM. }
\label{MDBMDW}
\end{figure}

We are also interested to the effect from other parameters in Eq.(\ref{space}) on the predictions of BR($B^0_q\rightarrow l_1l_2$) in the MRSSM. Taking the central values in Eq.(\ref{space}) as default, we plot the predictions of BR($B^0_s\rightarrow \mu\tau$) with the variation of those parameters in FIG.\ref{LSTUD} and FIG.\ref{MDBMDW}. The blank area in subfigure represents the excluded region by the constraints in TABLE.\ref{constraints}. The results show that varying those parameters has very little effect on the prediction of BR($B^0_q\rightarrow l_1l_2$)  which takes values in a narrow region.

\section{Conclusions\label{sec4}}
In this paper, we analyze the LFV decays $B^0_q\rightarrow l_1l_2$ within the framework of the minimal R-symmetric supersymmetric SM, while considering the constraints imposed by experimental data on the parameter space. By scanning over 14 parameters as shown in TABLE.\ref{scan}, we obtain a set of values which can reproduce the SM-like higgs mass in the MRSSM at $1\sigma$ level. Within this parameter space, we show that the prediction on BR($B^0_q\rightarrow l_1l_2$) depends strongly on the mass insertion parameters $\delta^{ij}_L$.  The logarithm base 10 of $\delta^{ij}_L$ are constrained to be log$_{10}\,\delta^{12}_L\leq -2.6$ , log$_{10}\,\delta^{13}_L\leq - 0.2$ and log$_{10}\,\delta^{23}_L\leq  - 0.1$ by radiative two body decays $l_2\rightarrow l_1\gamma$ respectively. We also show that the prediction on BR($B^0_q\rightarrow l_1l_2$) depends strongly on $\tan \beta$ and the mass insertion parameters $\delta^{ij}_Q$. The upper prediction on BR($B^0_q\rightarrow e\mu$) is around $\mathcal{O}(10^{-18})$. The upper prediction on BR($B^0_q\rightarrow l_1\tau$) is around $\mathcal{O}(10^{-10})$. In particular, the upper prediction on BR($B^0_d\rightarrow \mu\tau$) is four orders of magnitude below future experimental limit and we may make more efforts to observe it in future experiment. 

\begin{acknowledgments}
\indent\indent
This work has been supported partly by the National Natural Science Foundation of China (NNSFC) under Grant No. 11905002, the Natural Science Foundation of Hebei Province under Grants Nos. A2022104001, the Natural Science Basic Research Program of Shaanxi (Program No. 2023-JC-YB-072), and the Foundation of Baoding University under Grant No. 2023Z01.

\end{acknowledgments}

\end{document}